\documentclass[aip]{revtex4-1}
\usepackage{graphicx}
\usepackage{bm}
\draft 
\newcommand{\be}{\begin{equation}}
\newcommand{\ee}{\end{equation}}
\newcommand{\eea}{\end{eqnarray}}
\newcommand{\bea}{\begin{eqnarray}}

\begin{document}


\title{A Lie Scale Invariance in Fluids with Applications } 



\author{R. N. Henriksen}
\affiliation{Physics, Engineering Physics, Queen's University at\\Kingston, Ontario, K7L 3N6, Canada.}


\date{\today}

\begin{abstract}
 Lie scale invariance is used to reduce the incompressible Navier-Stokes equations to non-linear ordinary equations. This yields a formulation in terms of logarithmic spirals as independent variables. We give the equations when the spirals lie on cones as well as in planes. The theory gives a locus in cylindrical coordinates of singularities as they arise in the reduced Navier-Stokes equations. We give two formal examples aimed at discovering singularities in the flow; another example is related to a Hele-Shaw cell, and finally we explore the flow through propellers comprised of blades made from congruent logarithmic spirals.  
 
\end{abstract}


\maketitle 

\section{Introduction}
\label{sect:intro}
The implications of the Navier-Stokes equations as a function of scale is a famous problem. Applications normally appear in the theory of turbulence \cite{MK2000},\cite{USR2019}. One way of proceeding is to translate these equations into a self-similar, that is a scale independent form. We do this  following the Lie formalism  described in Carter and Henriksen\cite{CH1991} and elaborated in Henriksen\cite{H2015}. 

The number of exact solutions (e.g.\citep{DR2006}) is increased by our study and the corresponding Euler example contains singularities (e.g. \citep{MB2001}), although one of which is `built in' by the $1/\delta r$ scaling.

 Our technique uses the physical symmetry dictated by the scaling involved in dimensional analysis. The physical dimensions of the problem are assembled into a dimensional Lie scaling vector ${\bf a}$. In fluid mechanics this becomes ${\bf a}=(\delta,\alpha,\mu)$, which components correspond respectively to scaling in space, time and mass. Each physical quantity that appears in the problem must also be assigned a numerical "dimensional covector" ${\bf d}$ in the scaling space of ${\bf a}$. Thus for example velocity has the covector 
${\bf d}_v=(1,-1,0)$ and density has the covector ${\bf d}_\rho=(-3,0,1)$. 

The symmetry is applied by defining the increment $dT$ along the chosen scaling direction. In the present case we choose this to be the radial direction in cylindrical symmetry. The scaling symmetry requires each physical quantity $\Psi$ to be expressed as
\be
\Psi={\overline\Psi}~e^{({\bf d}_\Psi\cdot {\bf a})T},\label{eq:symmetryeq}
\ee
where $\overline\Psi$  must be invariant under the Lie motion, that is independent of $T$. The variables on which $\overline\Psi$ depends must therefore also be independent of $T$. Below, we use the scaling invariant angle $\Phi=\phi-\epsilon \delta T$ and the scaling invariant ratio $\zeta=z/r$ for the steady state variables. When time is present,the true scaling symmetry requires the addition of the invariant $\tau=te^{\alpha T}$. 

Because the radius is the scaling direction here, we must have
$\delta r=e^{\delta T}$ where $\delta$ is kept as the explicit spatial scaling factor and $T$ is a function only of $r$. According to the preceding general prescription, the velocity must take the form $v=u~e^{(\delta-\alpha)T}$, and the density becomes $\rho=\overline\rho e^{(-3\delta+\mu)T}$. In incompressible flow the density is held constant under the Lie motion so that $\mu=3\delta$. The scaling space is thus reduced to ${\bf a}=(\delta,\alpha)$. Moreover, $p/\rho=H e^{(2\delta-2\alpha)T}$ where $H$ is invariant.

This procedure is a multi-variable self-similarity analysis that yields a set of equations for the physical scale invariants ${\bf u}$ and $H$ as a function of $\Phi$, $\zeta$ and $\tau$. The problem is reduced by one independent variable $r$ by the application of the symmetry. However, this is still cumbersome and we use a single combination of the invariants in this paper.

In this paper we restrict ourselves to incompressible, viscous flow, which has a viscosity largely determined by the scale invariant requirement. 
The Lie symmetry is taken to be in the radial direction in this paper. An alternate, but parallel, discussion would take the Lie symmetry to be in the axial $(z)$ direction.

Our scale invariant variable represents an expansion of the flow in terms of equiangular (i.e logarithmic) spirals at different times. A flow of this type, free of controlling surfaces, would be subject to a two point boundary condition in angle. With external controls, a boundary condition would be applied on one of these spirals made into a real surface. In such a case our study with radial inflow has some resemblance to the inverse sprinkler problem \cite{WSZR2024}. We use it here to investigate the locus of singularities in cylindrical coordinates, to study flat Hele-Shaw type flows, and equi-angular propellers. The latter example requires the logarithmic spirals to lie on cones rather than in a plane. This will be our major example.

The cylindrically symmetric, time dependent, limit (when the spirals are circles) may be solved analytically, as can a related model in the Euler equations. The singularity structure, or the lack of it, may be seen at all scales in these examples.  These contribute to the limited number of known analytic solutions to the Navier-Stokes equations (\cite{DR2006}) both without and with singularities.

This is a purely formal result, but a very approximate conceptual model of such flow is shallow water exiting through a circular pipe out of a surrounding circular reservoir. The fluid must have been given some initial angular momentum about the pipe axis. We have added angular dependence in the Euler example without viscosity. 


In the next section we apply the technique. The resultant formulation of the Navier-Stokes equations is the principal result of this paper. The subsequent sections describe examples of some physical interest, and also of some mathematical interest regarding singularities in the equations.

\section{Scale Invariant Variables}
\label{sect:self-similar}

 By `scale invariant variables'  we refer particularly to the formalism introduced in \cite{CH1991} and at length in \cite{H2015}.

We take the Navier-Stokes equations to be  
\bea
\nabla\Psi+\partial_t({\bf v})&=&{\bf v}\times{\bf\omega}-\nu\nabla\times{\bf\omega},\nonumber\\
\nabla\cdot{\bf v}&=&0\label{eq:N-S}
\eea
where the vorticity and the sum of the thermal, kinetic and potential energies are                         
\be
\omega\equiv \nabla\times {\bf v},~~~~~~~~\Psi=p/\rho+{\bf v}^2/2 +V,\label{eq:defs}
\ee

The external potential $V$ can be included only if it happens to fit the scale invariant functional form. The fluid velocity is represented by ${\bf v}$, the pressure is $p$, the constant density is $\rho$ and the kinematic viscosity is $\nu$.

In this paper we use a Lie symmetry in radius (we refer to a cylindrical coordinate system $(r,\phi, z)$), according to 
\be
\delta r=e^{\delta T},\label{eq:Lie1}
\ee
because, for any length, the dimensional co-vector is ${\bf d}=(1,0)$\footnote{Here the co-vector ${\bf d}$ is resolved along space-time units so for the length co-vector, there is one unit of space and zero units of time}. The parameter $T$ depends only on $r$. The quantity $\delta$ is a reciprocal length scale that can be chosen at will to study different scales in the flow.  

Along this radial scaling motion we must have (the angle is discussed below) for complete consistency 
\bea
\delta z&=&\zeta e^{\delta T},\nonumber\\
t&=&\tau e^{\alpha T},\label{eq:indepvars}
\eea
where \cite{H2015} $\zeta$ and $\tau$ are invariant under the scale change. That is they can not depend on $T$ (i.e. $r$), but rather only on some combination of variables that is also invariant. The quantity $\alpha$ is another reciprocal length scale dedicated to the Lie scaling in time. 

For example according to equations (\ref{eq:Lie1}) and (\ref{eq:indepvars}), $\zeta=z/r$. The total derivative of $\zeta$ with respect to $r$ must be zero which requires $dz/dr=\zeta$. 

We do not allow a full variable $\zeta$ dependence in this paper, because this would double the independent variables. It appears in our equations through a scale invariant variable $\xi$ (see e.g. equation (\ref{eq:comvariable} below), but $\zeta$ has to be set subsequently to an arbitrary constant value in the equations to avoid breaking the symmetry
. 
We shall see below that the variable $\xi$, when set equal to a constant, describes a conical surface bearing  equi-angular spirals, where the exterior conical angle $\theta$ is given by  $ \tan(\theta)=dz/dr=\zeta$.

Allowing the local time to scale with radius as indicated in equation (\ref{eq:indepvars}), also breaks the radial Lie symmetry by introducing $\tau$ into the equations. Holding $\tau$ constant (as we do with $\zeta$ when present) would imply looking at different radii at different times, just as holding $\zeta$ constant implies different $z$ at different radii.  
 Instead, we can replace the local time scaling in equation (\ref{eq:indepvars}) with a global time scaling that is independent of radius or azimuth. We give this global definition of $\tau$ below. This  strictly contradicts the assumed Lie symmetry. However it turns out that in one particular case ($\alpha=0$), the equations admit a global, intrinsic, time scale. For the Euler equations, the intrinsic time scale is replaced by an external time scale but $\alpha$ remains zero. The more general case, classified by $a\equiv \alpha/\delta$  is restricted to the steady state. 

We keep the arbitrary reciprocal scales $\delta,\alpha$ explicitly. In effect they vary our resolution of the flow as a kind of mathematical `loupe'.
It happens that scale invariant solutions of mechanical systems  depend on the ratio of the reciprocal spatio/temporal scales, which ratio we designate by  
\be
a=\alpha/\delta,\label{eq:a}
\ee
and which is elsewhere referred to as the similarity class  \cite{CH1991}$^,$ \cite{H2015}.

The scale invariant angle (i.e. the total radial derivative of $\Phi$ is constant) is defined \cite{CH1991} as ($\epsilon T$ is an arbitrary rotation rate along the Lie scaling direction)
\be
\Phi=\phi-\epsilon T,\label{eq:Phi}
\ee
but we will use it multiplied by an arbitrary constant $k_\phi$ in the form
\be
\Phi=k_\phi\phi+k_r\delta T.\label{eq:Phivar}
\ee
Here, $k_r$ is another arbitrary constant that absorbs $\epsilon$.  
 
The same scaling symmetry is applied to the dependent quantities ${\bf v}$, $\bm{\omega}$ and the kinematic viscosity $\nu$, we follow the dimensional prescription for the variation along the Lie motion in $r$ to write them as 
\be
{\bf v}={\bf u}(\xi) e^{(\delta-\alpha)T},~~~~\bm{\omega}=\bm{\varpi}(\xi) e^{(-\alpha T)} ~~~~\nu=\nu_o(\xi) e^{(2\delta-\alpha)T},\label{eq:depvars}
\ee
where ${\bf u}$, $\bm{\varpi}$ and $\nu_o$ are now scale invariant quantities. The quantity $\xi$ is a place holder for the combined scale invariant variable that we define below.  Equations (\ref{eq:N-S}) are to be expressed in terms of these quantities. 

Technically, although this is not the subject of this paper, we have used use variable reciprocal scales, $\delta$  and $\alpha$ to form the space-time reciprocal dimensional vector (corresponding  in position to space and time) ${\bf A}=(\delta,\alpha)$. This forms a dot product with a (quantity dependent) numerical co-vector ${\bf d}$, which corresponds in position and sign to the number of spatial and temporal dimensions of the quantity of interest. The dot product gives the factor by which a quantity changes exponentially along a Lie scaling motion \cite{CH1991}),\cite{H2015}, namely $e^{({\bf A}\cdot{\bf d})T}$. Here the differential $dT$ (not to be confused with ${\bf d}$) is an increment along the Lie symmetry direction. Thus we have ${\bf v}=e^{(\delta-\alpha)T}{\bf u(\xi)}$, because $\xi$ is constant along the Lie motion.

Similarly 
\be
p/\rho=H(\xi)e^{2(\delta-\alpha)T}~~~~V=W(\xi)e^{2(\delta-\alpha)T}.\label{eq:auxilliary}
\ee

The expression for $p/\rho$  is naturally part of the description of an incompressible, scaled fluid; but the scaling form for the external potential is a severe constraint. Subsequently we will not include the potential explicitly; but it can be included through $W(\xi)$ summed with $H(\xi)$, if it is compatible with the assumed symmetry.


Provided that the kinematic fluid viscosity in equation (\ref{eq:depvars}) is constrained by requiring $\nu_o$ to be a global constant (i.e. independent of $\xi$), we can define a global scaled time as
\be
k_o\tau=k_o(\delta^2\nu_o)t.\label{eq:tau}
\ee
Here $k_o$ is a dimensionless constant, whose value allows some external forcing. For the Euler equations, we will use $\tau=k_o t$ where $k_o$ is an externally applied reciprocal time.
The spatial scaling implied by the constant $\delta$, allows the viscous global time scale to vary with an arbitrary length scale. The fixed time $k_o\delta^2\nu_o$ at any spatial scale removes any additional temporal scaling of the system \cite{CH1991}. Hence we will require $\alpha=0$ (and hence $a=0$), whenever the flow is not steady.

The three independent scale invariant quantities $\zeta$,$\tau$, and $\Phi$ can be used in principle to express equations (\ref{eq:N-S}), rather than using $(t,\phi,z)$. This removes $r$ from the independent variables \cite{CH1991}, but leaves a multi-variable problem even in the steady state.

To proceed more simply, we look for solutions of a single combined scale invariant variable, $\xi$. We choose a sum of the individual invariants $\Phi$, $\zeta$ and $\tau$. In principle $\xi$ might be any function of the scale invariants, but such a choice is not guaranteed to eliminate all of coordinates from the equations.

The sum of $\Phi$,$\zeta$ and $\tau$ defines a spiral basis through (we use the global definition of $\tau=\delta^2\nu_ot$) 
\be
\xi=k_\phi\phi+k_r\delta T+k_z\zeta+k_o\tau.\label{eq:comvariable}
\ee
The quantity $k_o$ allows for some external forcing instead of the viscous decay.
In the Euler limit, the time dependence is simply $k_o t$, where $k_o$ has the dimension of reciprocal time and might be an applied frequency.

We allow the constants $\{k_r,k_\phi\}$ (and when applicable $k_z$ and $k_o$) to be positive or negative. In general, the solutions are single valued only over an interval of the combined variable $\xi$.

Although we do not consider a self-gravitating fluid in this paper it is useful to consider such an example to show the physical significance of the index $a$.
We assume for this purpose an axially symmetric, steady, ($k_\phi=k_o=0$) cylindrical distribution of mass. We take a power law (scale invariant) density distribution $\rho=\lambda/(\delta r)^q$. This produces a self gravitational potential ($q>2$ for a bound system)
\be
V=-\frac{4\pi G\lambda}{\delta^2 (2-q)^2}(\delta r)^{(2-q)},
\ee
which for compatibility with the scale invariant form should equal  
\be
W(\xi)e^{2(1-a)\delta T}=W(\xi)(\delta r)^{2(1-a)}.
\ee
Consequently, we must have $W$ to be constant and $a=q/2$. When $q=2$, and there is no line mass at the centre, we require $a=1$ and $W(\xi)=-(2\pi G\lambda/\delta^2) \xi^2 $ with $k_r=1$ and $k_\phi=k_o=0$. Note that $\xi=\ln{(\delta r)}$ in this case.

Despite this illustrative toy example; in this paper, $q=0$ for incompressibility, and an applied potential is not explicitly discussed. The index $a$ is therefore not constrained in a steady state.

In the next section we reduce the Navier-Stokes equations (\ref{eq:N-S}) to non linear ordinary equations in terms of the scale invariant variable.  As far as we know this particular formulation is original (although see e.g.  \cite{H2015}).Subsequent sections will explore special cases  without being exhaustive.

\section{Navier-Stokes Equations as Ordinary Equations}
\label{sect:equations}

\subsection{Steady State}
\label{sect:steadystate}

We substitute the expressions (\ref{eq:depvars}), (\ref{eq:indepvars}) and (\ref{eq:auxilliary}) together with the definition of $\xi$ (\ref{eq:comvariable}) into equations (\ref{eq:N-S}). We write the steady equations with $k_o=0$ and keep the similarity class $a$ general. We find (the prime indicates the derivative with respect to $\xi$ as defined in equation(\ref{eq:xi1})) the following scale invariant equations;

The incompressible condition,
\be
(2-a)u_r+k_r u_r'+k_\phi u_\phi'+k_z u_z'-k_z\zeta u_r'=0\label{eq:div1};
\ee
When $k_z\ne 0$, our equations apply only where $\zeta=constant$. In strict cylindrical symmetry $k_z=0$. In the steady state the independent variable $\xi$(see \ref{eq:comvariable}) is,
\be
\xi=k_\phi \phi+k_r\ln{\delta r}+k_z\zeta.\label{eq:xi1}
\ee

It is convenient to introduce a velocity along the ${\bf k}$ vector (perpendicular to the surface of constant $\xi$) as
\be
u_k\equiv{\bf k}\cdot{\bf u}=k_ru_r+k_\phi u_\phi+k_z u_z,\label{eq:uk}
\ee
and 
\be 
k^2=k_r^2+k_\phi^2+k_z^2.\label{eq:k}
\ee
We may use the continuity equation to eliminate $u_r$ in favor of $u_k$ whenever $k_z$ and $\zeta=0$
\be
(2-a)u_r=-u_k'.\label{eq:div2}
\ee
The azimuthal velocity is then given in terms of $u_k$, $u_r$ and $u_z$ by
\be
k_\phi u_\phi= u_k-k_ru_r.\label{eq:uphik}
\ee
These expressions are also simplified if $a=2$ (when viscosity is uniform), because then  equation (\ref{eq:div2}) can be integrated.

The radial component of the Navier-Stokes equation becomes (we show an example of the necessary algebra in an Appendix),

\bea
&2(1-a)(H+\frac{{\bf u}^2}{2})+(k_r-k_z \zeta) H'=(2-a)u_\phi^2-u_k u_r'+(1-a)u_z^2+k_z\zeta u_r u_r'\nonumber\\
&-\delta\nu_o [(2-a)k_\phi u_\phi'-k^2 u_r''+k_r u_k'' -a k_z u_z']+ \delta\nu_o~ k_z ~\zeta~(k_\phi u_\phi''+k_z u_z'');\label{eq:NSradial}
\eea

The azimuthal component of the Navier-Stokes equation,
\bea
k_\phi H'&=&-u_k u_\phi'-(2-a)u_r u_\phi+k_z \zeta u_r u_\phi' +\delta\nu_o[-a(2-a)u_\phi\nonumber\\
&+&2(1-a)k_ru_\phi'+k^2u_\phi''-k_\phi u_k''+a k_\phi u_r']\nonumber\\
&-&~\zeta ~\delta\nu_o~\big((1-2a) k_z u_\phi'+k_z(2k_r-k_z \zeta)u_\phi''-k_z k_\phi u_r''\big);\label{eq:NSazimuth}
\eea

The longitudinal component of the Navier-Stokes equation,
\bea
k_z H'&=&-u_r (1-a)u_z-u_k u_z'+k_z \zeta u_r u_z'\nonumber\\
&+&\delta\nu_o[(1-a)((1-a)u_z+2k_r u_z')+k^2u_z''-k_z u_k''-k_z u_r'(1-a)]\nonumber\\
&-&\delta\nu_o\big((1-2a)k_z \zeta u_z'-k_z^2\zeta u_r''+k_r^2u_z''-(k_r-\zeta k_z)^2 u_z''\big).\label{eq:NSvertical}
\eea

Should $a=2$, the viscosity is constant in space and time, which is the most common assumption\cite{F1995}. The value $a=1$ implies that velocities (and energies) are constant under the Lie motion in radius. It is usually termed `homothetic' in general relativity, where it implies that all scales are changed by the same factor. 

Other values of $a$ correspond to constants with other dimensions. For example $a=1/2$, that is $\delta=2\alpha$, corresponds to a constant acceleration under the Lie motion in radius. The value $a=3/2$ implies a velocity that scales like $L^{-1/2}$, while $a=4/3$ implies the Kolmogorov velocity scaling $\propto L^{-1/3}$. 

Appendix A gives the derivation of equation (\ref{eq:NSradial}) as an example of the necessary algebra.

\subsection{Time Dependence}
\label{sect:timedependence}

When $k_o\ne 0$ so that time dependence is possible, we must take $a=0$ and $k_o\ne 0$. The Lie invariant becomes ( recall $\delta T=\ln{\delta r}$ and $\tau=\delta^2\nu_o t$)
\be
\xi=k_\phi\phi+k_r\delta T+k_z\zeta+k_o\tau.\label{eq:xi2}
\ee
This gives the flow equations (plus the continuity equation \ref{eq:div1}) as:

Radial equation;
\bea
&2&H+(k_r-k_z \zeta) H'+k_o\delta\nu_o u_r'= u_\phi^2-u_k u_r'-u_r^2 +k_z\zeta u_r u_r'\nonumber\\
&-&\delta\nu_o [2k_\phi u_\phi'+k_r u_k''-k^2 u_r'']+\delta\nu_o k_z\zeta~(k_\phi u_\phi''+k_z u_z'');\label{eq:NSradialt}
\eea

Azimuthal equation;
\bea
&k_\phi H'+k_o \delta\nu_o u_\phi'=-2u_r u_\phi-u_k u_\phi'+k_z \zeta u_r u_\phi'\nonumber\\
&+\delta\nu_o(2k_r u_\phi'+k^2 u_\phi''-k_\phi u_k'')]-\delta\nu_o~\zeta~\big((1-2a)k_z u_\phi'+k_z(2k_r-k_z \zeta)u_\phi''-k_z k_\phi u_r''\big); \label{eq:NSazimutht}
\eea

Longitudinal equation;
\bea
&k_z H'+k_o\delta\nu_o u_z'=-u_r u_z-u_k u_z' +k_z\zeta u_r u_z'+\delta\nu_o[u_z+2k_r u_z'\nonumber\\
&+k^2 u_z''-k_z u_k''-k_z u_r']+\delta\nu_o\big(k_z^2 \zeta u_r''-k_z \zeta u_z'-k_r^2u_z''+(k_r-k_z \zeta)^2u_z''\big);\label{eq:NSverticalt}.
\eea
For completeness we proceed next to the zero viscosity Euler equations under this formulation.

\subsection{Euler Equations}
\label{sect:Eulereqs}

We obtain the Euler equations under this symmetry by setting $\nu_o=0$ in the equations of the preceding steady state  section, except on the left side of each equation when there is time dependence. In that case we replace $\nu_o$ on the left of each time dependent equation with $1/\delta^2$, and $\tau=k_o t$ in the variable $\xi$, so that $k_o$ has the dimension of reciprocal time. 

The Euler equations become explicitly (plus equation \ref{eq:div1}) for the steady state:

Radial equation;
\be
2(1-a)H+(k_r-k_z\,\zeta) H'=u_\phi^2-u_ku_r'-(1-a)u_r^2 +k_z \zeta u_r u_r',
\ee

Azimuth equation;
\be
k_\phi H'=-u_k u_\phi'-(2-a)u_r\,u_\phi +k_z \zeta u_r u_\phi',
\ee

Longitudinal equation;
\be
k_z\,H'=-(1-a)u_r\,u_z-u_k\,u_z' +k_z \zeta u_r u_z'.
\ee
The continuity equation is unchanged from equation (\ref{eq:div1}). The independent variable is $\xi=k_\phi\,\phi+k_z\,\zeta+k_r\,\delta T$, where $\delta T=\ln{\delta\,r}$. 

The Euler equations with time dependence are as above in this section but with $a=0$. There is also an extra term $(k_o/\delta)\, u_x'$ on the left of each equation, where $u_x'$ indicates the derivative of the corresponding velocity ( i.e. $u_r'$, or $u_\phi'$, or $u_z'$).

\section {Thick Disc with Time Dependence}
\label{sect:axiallysymmetric}

In this section we study an axially symmetric disc in viscous flow. The solution is expressible either in terms of elementary functions or as  well studied confluent hyper geometric functions.  The disc may have an arbitrary thickness, but each section of constant $z$ behaves in the same way. 

This problem is of  some astrophysical interest (although magnetic fields have replaced viscosity currently, our technique could generalize to include magnetic fields) and was studied in a classic paper (\cite{LBP1974}), which allows some comparison with this introductory example. The main difference is our choice of the Lie similarity symmetry, which requires the viscosity to increase with radius squared. Moreover we have simplified the physical arguments by taking constant spatial density in the flow. A variable surface density might be calculated as $\sigma(r)=\int~v_z(t,r)) dt$, but we will not dwell on the detailed application in our example.  

An astrophysical disc will have both an inner and an outer radial boundary, corresponding to the inner protostar and an outer cloud source.  
A purely fluid illustration is water flowing with rotation down the sink in a flat bottomed wash basin.

The mathematical description follows from the equations of section  (\ref{sect:timedependence}) by setting $k_\phi=0$, $k_z=0$ and $a=0$. Setting also $\zeta=dz/dr=0$ defines the disc behaviour. 

From equation (\ref{eq:div1}) we find the scaled radial velocity as
\be
u_r=V_r\exp{(-\frac{2}{k_r}\xi)},\label{eq:conax}
\ee
where 
\be
\xi=k_r\ln{\delta r}+k_o\delta^2\nu_o t,\label{eq:axkappa}
\ee
and $V_r$ is a constant. 
 We recall that  
\be
{\bf v}=e^{\delta T}{\bf u}\equiv (\delta r){\bf u},\label{eq:velocity} 
\ee
so that 
\be
v_r=\frac{V_r}{(\delta r)} e^{-(\frac{2k_o\delta^2\nu_o t}{k_r})},\label{eq:vr1}
\ee
on using equations (\ref{eq:conax}) and (\ref{eq:axkappa}). At $t=0$ and $\delta r=1$ we have $V_r=v_r(0,r)$ as our initial condition.

The constants  $k_o$, $k_r$ are determined either by external forcing or by a two point eigenvalue problem, should the latter be relevant between two fixed radii.

We note that if we follow constant radial velocity as we vary the scale then $\delta r$ and $\delta^2 t$ are constant. Hence  $r\propto 1/\delta$ and $t\propto 1/\delta^2$ so that $r^2/t$ is constant for the same velocity at different scales. In time, $\delta r$ must decline proportionately to the exponential factor. 

The viscosity is, with $a=0$,
\be
\nu=\nu_o (\delta r)^2,\label{eq:viscax}
\ee
and so is constant with constant $\delta r$. Hence the $r^2/t$ behaviour is to be expected with scale. In reality, such a radial variation of viscosity at fixed scale would be due to an appropriate temperature decline of the fluid or, perhaps to an appropriate fluid mixture.

The inward or outward  mass flow per unit length along the vertical axis is  constant in space at fixed $\delta$, although not in time. Because the divergence is zero, this flux must be redirected along the $z$ axis if the flow reaches the axis. In an astrophysical context the flow would be accreted by the protostar.  This may require a radial two point boundary solution, if something like the protostar luminosity were known.

The only driving potential compatible with the scale invariance with $a=0$ has the harmonic form $W(\xi)(\delta r)^2$, where $W$ may be taken constant (\ref{eq:auxilliary}). It corresponds approximately at the central section to the gravitational potential of a short column of constant density ($W=\pi G\rho/\delta^2$).  It would be included automatically in $H$.
 
We continue the example by examining the equations for the other components of the flow based on the formulation given in section (\ref{sect:timedependence}).

Equation (\ref{eq:NSradialt}) for the scaled radial velocity becomes
\be
2H+k_r H'=u_\phi^2-u_r^2-(k_ru_r+k_o\delta \nu_o)u_r',\label{eq:rax}
\ee
which is an equation for the variation of $H(\xi)$ once $u_\phi$ is known. 

 Equation (\ref{eq:NSazimutht}) for the scaled azimuthal velocity becomes a homogeneous linear equation  
\be
k_r^2u_\phi''-(k_o-2k_r+\frac{k_r u_r}{\delta \nu_o})u_\phi'-2(\frac{u_r}{\delta\nu_o})u_\phi=0,\label{eq:azax}
\ee
which can be solved formally together with equation (\ref{eq:conax}). We ignore the relatively uninteresting case with $u_\phi=0$. 
We set $V_r/\delta\nu_o=-u$ in equation (\ref{eq:azax}) so that $u$ is positive for inward flow. 
The solution of equation (\ref{eq:azax}) takes the form (the outflow case with $u=+1$ must be found separately)
\bea
&u_\phi&=C_1(-k_o+k_r ue^{-2\xi/k_r})\nonumber\\
&+&C_2~ M\big([-k_o/2k_r],[2-k_o/2k_r],(u/2)e^{-2\xi/k_r}\big)\exp{((k_o/k_r^2-2/k_r)\xi)},\label{eq:uphia0ax}
\eea
where $M(a,b,S)$ represents the single valued confluent hypergeometric function (e.g. \cite{abst1970}). 
The function $M(a,b,S)$ behaves smoothly without singularity unless $b=-m$ and $a=-n$ with $m<n$ where $m$,$n$ are integers.
In that case $M$ is undefined and the solution of equation (\ref{eq:azax}) must be found numerically. Such solutions also show no point singularity.

If we recall that $v_\phi=\delta r u_\phi$ and that $\exp{(-2/k_r\xi)}=e^{-2(k_o/k_r)\tau}/(\delta r)^2$, then the $C_1$ solution always asymptotes in time ($k_o/k_r>0$) to rigid body rotation so that angular momentum increases outward. At $t=0$ there was a linear combination of rigid rotation and conserved angular momentum.  

The $C_2$ solution also asymptotes to rigid body rotation provided $k_o/2k_r=1$, because $M(a,b,z)\rightarrow constant$ as $z\rightarrow 0$, which it does in increasing time. If $k_o/2k_r<1$ this solution asymptotes to zero in time and, if $k_o/2k_r>1$, this solution diverges in time. For both solutions to obey the energy argument yielding asymptotic rigid rotation (\cite{LBP1974}), we should use $k_o/2k_r=1$.

The vertical equation for $u_z$ from equation(\ref{eq:NSverticalt}) is the homogeneous linear equation 
\be
k_r^2u_z''-(k_o-2k_r+\frac{k_ru_r}{\delta\nu_o})u_z'+(1-\frac{u_r}{\delta\nu_o})u_z=0.\label{eq:verax}
\ee
We will set $V_r/(\delta\nu_o)=-u$ to correspond to the solution for the azimuthal velocity.

The formal solution of the linear equation (\ref{eq:verax}) may also be found in terms of conformal hypergeometric  functions $M$ and $U$  as 
\be
u_z=(C_1~ M(a,b,S)+C_2~U(a,b,S))e^{w\xi}.\label{eq:uzgen}
\ee
The defined quantities are
\bea
a&\equiv&(1/4)(\sqrt{(k_o^2/k_r^2-4 k_o/k_r)}-(k_o/k_r)),~~~~b\equiv1+(1/2)(\sqrt{(k_o^2/k_r^2-4 k_o/k_r)})\nonumber\\
S&\equiv&(u/2)e^{-2\xi/k_r},~~~~w\equiv (1/2k_r^2)((k_o-2k_r)-\sqrt{(k_o^2-4 k_r k_o)}).\label{eq:uzdetails}
\eea

In this case the parameters $a$ and $b$ can be complex and so the real part of the hypergeometric functions must be taken. 

The interesting result is that with $k_o=2$, $k_r=1$ ,both $M$ and $U$ yield $u_z$ oscillating as $\xi$ 
 runs over positive values. The period is $\Delta\xi\approx 6$. This oscillation will happen in time at fixed $\delta r$, but at fixed time it is smoothed by the logarithmic dependence on radius. The KummerM dependence yields a large  $u_z$ at small $\xi$ ($\xi\approx 0.2$) and at the positive or negative peak of each oscillation. 
Asymptotic forms are available for both large and small $z$, although the principal branch of KummerU lies in the range $-\pi<z<\pi$.

 If one wanted a two sided thick disc, the zero $u_z$ solution
 would apply at $z=0$, followed by the negative $u_z$ solution at negative $z$. The oscillations appear to be due to a strong peak in the pressure function at small $\xi$.

\section{Flat Fluid Sheet}
\label{sect:Hele-Shaw}

In contrast to the preceding section, we study the motion and instability of a flat sheet of water obeying our Lie scaling symmetry. We use cylindrical coordinates with $z$ perpendicular. Time dependence and non axial symmetry are included, so that we must set $a=0$ as before. We will study the case with $u_z=0$, and hence also $k_z=0$ for consistency with the longitudinal equation of section (\ref{sect:timedependence}).  Because of the sheet constraint we also set $\zeta\equiv dz/dr=0$, which also implies $z=0$ by definition. 

The sheet has, according to our basic symmetry (\ref{eq:depvars}), a viscosity increasing in radius as $(\delta r)^2$, and our example might be considered as an approximation to a Hele-Shaw cell (e.g.\cite{GVN2024}). However, although setting $u_z=0$ can allow for flat boundaries above and below the $r$, $\phi$ plane, the no-slip condition can not be applied without a $z$ dependence. Moreover, the viscous variation is smooth rather than discontinuous. Nevertheless, there is some similarity in behaviour.

When $a=0$ and $\zeta=0$ we have from the equations of section (\ref{sect:timedependence}) that the planar velocities are 
\be
u_r=-\frac{u_k'}{2},~~~~~k_\phi u_\phi =u_k+\frac{k_r u_k'}{2},
\label{eq:planevels}
\ee
and we recall the variable
\be 
\xi=k_\phi\phi+k_r\ln{\delta r}+k_o\tau.\label{eq:xiHC}
\ee

The radial and azimuthal equations may now be expressed as two non linear equations for $u_k(\xi)$ and $H(\xi$. These are respectively 
\bea
2&H&+k_rH'=-\frac{k_r^2+k_\phi^2}{2}u_k'''+(k_o/2-2k_r)u_k''-2u_k'\\
&+&\frac{u_ku_k''}{2}+\frac{u_k^2}{k_\phi^2}+(\frac{k_r^2}{k_\phi^2}-1)\frac{(u_k')^2}{4}+\frac{k_r}{k_\phi^2}u_ku_k',\label{eq:rH-C}
\eea
and
\bea
k_\phi^2H'&=&\frac{k_r(k_r^2+k_\phi^2)}{2}u_k'''+(2 k_r^2-\frac{k_o k_r}{2})u_k''+(2k_r-k_o)u_k'\\
&-&\frac{k_r}{2}u_ku_k''+\frac{k_r}{2}(u_k')^2.\label{eq:phiH-C}
\eea
In these equations $u_k$ is in units of $\delta\nu_o$ and $H$ is in units of $(\delta\nu_o)^2$.

\begin{figure}{}
\begin{tabular}{cc}
\rotatebox{0}{\scalebox{0.5}
{\includegraphics{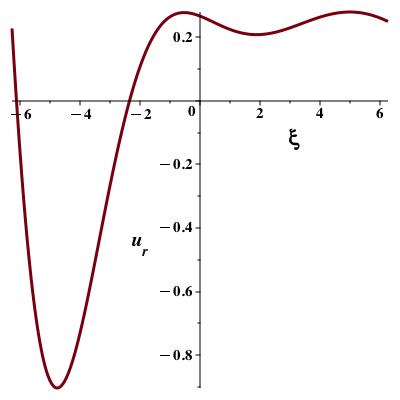}}}&
\rotatebox{0}{\scalebox{0.45}
{\includegraphics{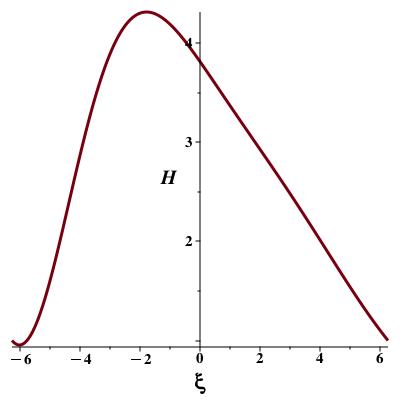}}}\\
\rotatebox{0}{\scalebox{0.45}
{\includegraphics{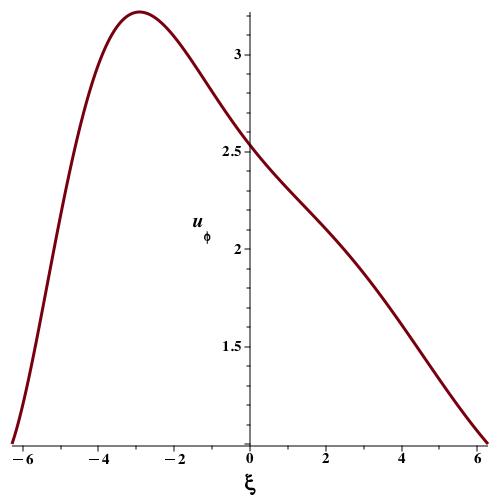}}}&
\rotatebox{0}{\scalebox{0.45}
{\includegraphics{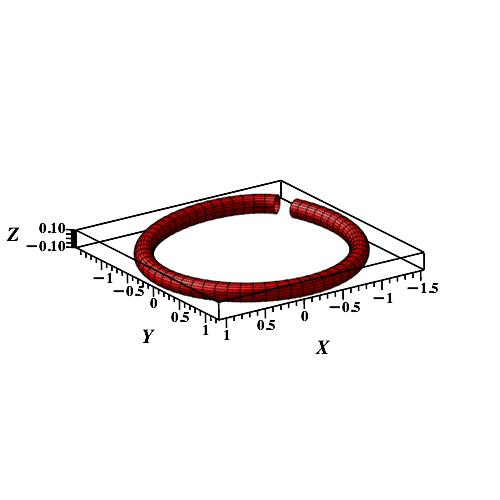}}}
\end{tabular}
\caption{ The figure shows the two point ($-6.28,+6.28$) flow velocities (to be multiplied by $r/\nu_o$, and one instantaneous 'stream line' cluster for the case when $k_\phi=2$,$k_r=1$ and the eigenvalue is $k_0=-1.91447$. The boundary conditions were $u_\phi=1$, $H=1$ at the end points, plus $u_k'(6.28)=-0.5$. The radial velocity is at upper left, the pressure function at upper right, the azimuthal velocity at lower left and a 'stream line' cluster is shown at lower right. The stream line cluster shown passes through $r=1$ and $\phi=0.1$ as a tube f radius $0.1$ centred on the plane $z=0$. \label{fig:H-S1}}\end{figure}

One way of determining the constants $k_o$, $k_\phi$ or $k_r$ is by solving  a two point boundary problem in either radius or angle, the end points corresponding to different $\xi$. If this is done say for $\Delta\phi=2k_\phi\pi$ at $\delta r=1$ and $\tau=0$, then the solution also holds on a ring moving according to 
\be
\delta r=e^{(-\frac{k_o\tau}{k_r})}.\label{eq:ring}
\ee
We show such an solution in figure (\ref{fig:H-S1}) using the two point numerical routine from MAPLE 2024. It  determines $k_o$ as an eigenvalue by requiring periodicity of the flow in $\xi$ over the range $(-2\pi, 2\pi)$. The end points were chosen rather than $(-\pi,\pi)$ because the remaining constants were chosen to be $k_\phi=2$ and $k_r=1$.

The value found for $k_o$ was $k_o=-1.91447$. Thus, according to equation (\ref{eq:ring}), the ring moves outward from $\delta r=1$ exponentially in time. The instantaneous stream line that comprises the ring illustrates a close confinement of the viscous fluid in this type of solution. We note that it is really only the axis of the ring in the plane that is relevant. The extention in $z$ is only for graphical visibility.

\clearpage

We turn to a more general type of solution where the fluid  flows more perceptibly. 
First we recall that the solution for the velocities (scaled by $\delta \nu_o$ is in terms of $\delta r$ (i.e.${\bf v}={\bf u}\delta r$) and $\xi$. Writing $\delta r$ as 
\be
\delta r=\exp{(\frac{\xi}{k_r}-\frac{k_\phi}{k_r}\phi-\frac{k_o}{k_r}\tau)},\label{eq:deltarphitau}
\ee
we see that a coordinate line of given $\xi$ will only arrive at the same radius, given a linear combination of azimuth and time equal to an appropriate constant. Different values of $\xi$ can not arrive at the same radius with the same combination of azimuth and time, so the planar lines of $\xi$ are good coordinates.  

\begin{figure}{}
\begin{tabular}{cc}
\rotatebox{0}{\scalebox{0.35}
{\includegraphics{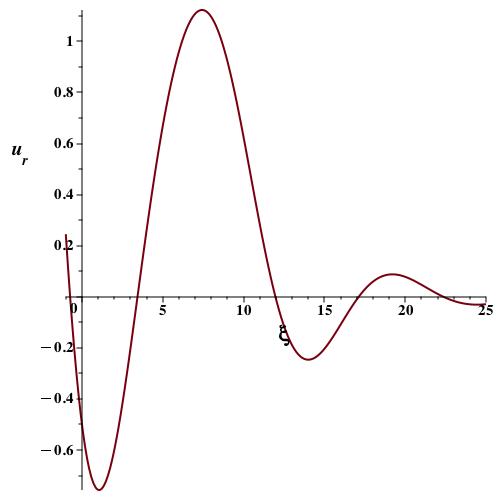}}}&
\rotatebox{0}{\scalebox{0.35}
{\includegraphics{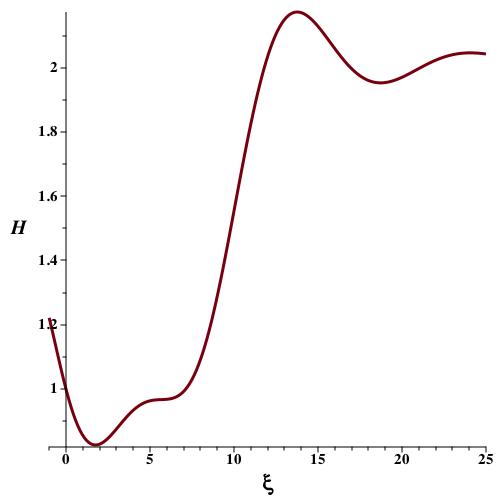}}}\\
\rotatebox{0}{\scalebox{0.35}
{\includegraphics{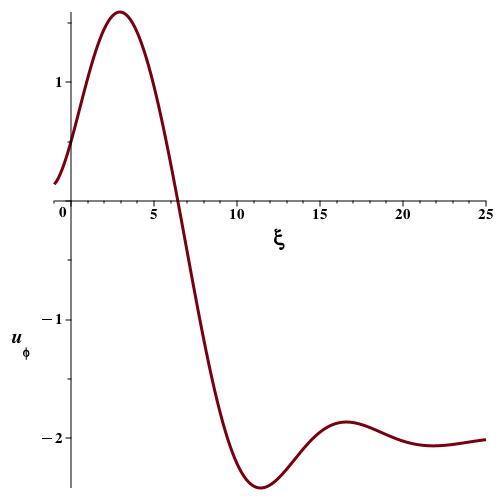}}}&
\rotatebox{0}{\scalebox{0.35}
{\includegraphics{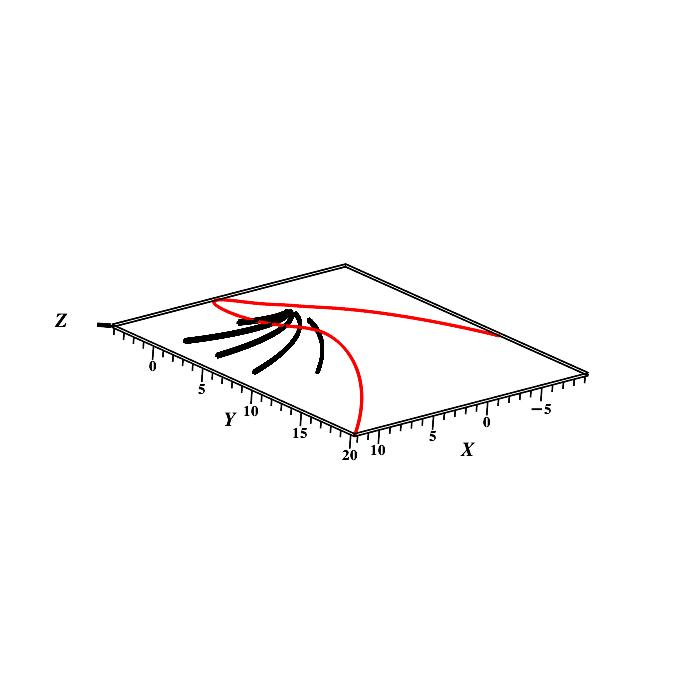}}}
\end{tabular}
\caption{ The figure shows the flow velocities as a function of $\xi$ and one instantaneous stream line ($\tau=1$) for the case when $k_\phi=3$,$k_r=1$ and $k_0$ is $0.5$.  The initial conditions were $u_k(0)=1$, $u_k(0)'=1$, $u_k''(0)=1$ and $H(0)=1$. The radial velocity is at upper left, the pressure function at upper right, the azimuthal velocity at lower left and one `stream line' is shown at lower right. For reference lines of constant $\xi$ are shown in the range $1.5\le 5.5$ in steps of $1$. They converge on $\delta r=1$ at different angles.  \label{fig:H-S2}}\end{figure}

In figure (\ref{fig:H-S2}) we show the flow velocities and the pressure function as functions of $\xi$, and a `stream line' (instantaneous at $\tau=1.0$) in the $xy$ plane. Five coordinate lines for $\xi>0$ are also shown for reference with the `stream line'. 

The instantaneous stream line passes through $r=3$ and $\phi=0.1$ (approximately $x=3$ and $y=0.3$). It may be regarded as entering on the $y$ axis at large $x$, and leaving on the $y$ axis at negative $x$ after a major bend.

The instantaneous `stream line' shown is not of course followed by any single fluid element, being comprised of an ensemble of elements at a fixed time. One can instead calculate the actual trajectory of an element that starts at a given time at a given point, but in fact such a trajectory is not qualitatively different from the ensemble, and the ensemble is more visually accessible in a highly viscous case ($ \nu_o$ large).

The question arises as to how an angular instability may be described in this formulation. Rewriting equation (\ref{eq:deltarphitau}) we see that at a fixed $\xi$ and $\phi$ The radius varies as 
\be
\frac{\delta r}{e^{-k_o\tau}}=\exp{(\xi-k_\phi\phi)},\label{eq:rxiphi}
\ee
so $\delta r$ increases/decreases exponentially in time with $k_o<0$/$k_o>0$. However, different values of $\xi$ and $\phi$ will increase at different rates, some faster and some slower. This implies  `fingering' \cite{GVN2024} towards the exterior. This expansion carries the flow velocities with it according to 
$(\delta r)(\delta\nu_o){\bf u}(\xi)$, and the pressure function with it according to $(\delta r)^2(\delta\nu_o)^2H(\xi)$.

To define the various fingers we expect a change in pressure radial gradient in $\xi$. This gradient has the same sign as the radial pressure gradient. A change in the sign of the of the radial pressure gradient happens at $\xi\approx 2$ and at $\xi\approx 14$ in figure (\ref{fig:H-S1}). However a choice of $\Delta\xi$ anywhere on a negative slope of $H(\xi)$ would also give a driving radial pressure gradient at the corresponding value of $\xi$. Resistance to the outflow corresponds to a positive slope of $H$. 

The angular extent of a `finger' in $\Delta\phi$ at  fixed $\delta r=1$ is  related to $\Delta\xi$ and $\Delta\tau$ by $\Delta \phi=(\Delta\xi-k_o\Delta\tau)/k_\phi$. Suppose we choose $\Delta\xi=1$ to encompass the dip in $H(\xi)$ at $\xi=\approx 2$. Then at fixed $t$, $\Delta\phi=1/3$ radian because $k_\phi=3$. This angular extent will shrink with increasing time until it grows again after reversing sign. This is a fingering instability.  The fingers are much narrower, the larger the value of $k_o$ or the smaller $\Delta \xi$. The finger will appear at different azimuth and time so long as $\xi$ is fixed in $\xi=k_\phi\phi +k_o\tau$.

\section{Propeller Geometry}
\label{sect:propeller}

We suggest the idea that surfaces comprised of a congruence of  $\xi$ lines near a common  value of $\xi$ (a continuous set of logarithmic spirals on a cone), could be useful as propeller blades. We study the fluid flow and a possible arrangement of the logarithmic spiral blades, but actual engineering of a logarithmic propeller must be done elsewhere.

We choose a steady state with $a=2$ so that viscosity of the fluid is constant and velocity (scaled by $\delta\nu_o$ is $\propto u/\delta r$. We take $\zeta=z/r=1$ for the tangent of the exterior cone angle, and hold $k_z$, $k_r$ and $k_\phi$ constant. Unlike the previous section we have no need of finding a two point solution because the flow will be bounded by `blades'. For simplicity we will explore the example with $k_r=k_z=1$ and $k_\phi=2$, but these are all variable in an engineering study.
A more restrictive assumption chosen purely for calculational  simplicity is
\be
k_r=1=k_z\zeta=1,\label{eq:kriskz}
\ee
but this can also be relaxed using the steady state equations.

The governing equations follow from section (\ref{sect:steadystate}) plus the continuity equation, when we apply the above assumptions. The problem is reduced (with our simplifying assumption) to solving for $u_\phi$ and $u_k$ because from continuity we have 
\be
u_r=\frac{u_k}{k_r}+u_{ro},\label{eq:urprop}
\ee
and from the definition of $u_k$ together with continuity and assumption (\ref{eq:kriskz})
\be
u_z=-\frac{k_\phi}{k_z} u_\phi-\zeta u_{ro}.\label{eq:uzprop}
\ee
In these relations $u_{ro}$ is an arbitrary constant.

The radial,integrated azimuthal, and longitudinal equations take the respective forms (dividing by $(\delta\nu_o)^2$)
\bea
-2H&=&u_r^2+u_\phi^2+u_{ro}u_k'+(k_\perp^2/k_r)u_k''-2k_\phi u_\phi',\\\nonumber
k_\phi H&=&k_ru_{ro}u_\phi+k_r u_\phi+(2k_\phi/k_r)u_k+k_\perp^2u_\phi'+k_\phi H_o,\\\nonumber
k_zH'&=& u_ru_z+k_ru_{ro}u_z'+k_ru_z'+k_\perp^2 u_z''+u_z+u_k'/\zeta.\label{eq:propeqs}
\eea
Here $H_o$ is a constant stemming from the integration of the azimuthal equation (which may also be used in differentiated form) and
\be
k_\perp^2=k_\phi^2+k_z^2.
\ee
The non linear Navier-Stokes equations for $u_\phi$ and $u_k$ as functions of $\xi$ become by combining the equations (\ref{eq:propeqs}) twice, to eliminate $H$ and  then assuming  $k_z=k_r/\zeta$;

\bea
(k_\perp^4/k_\phi)u_\phi'' &+&(k_r/k_\phi)(1+u_{ro})k_\perp^2 u_\phi'+(k_r/\zeta^2)u_k'+k_\phi(1+u_{ro})u_\phi\\\nonumber
&+&u_{ro}u_k+(k_\phi/k_r)u_\phi~u_k+k_ru_{ro}(1+u_{ro})=0,\label{eq:uphiprop}
\eea

and
\bea
&\frac{k_\phi^2+k_z^2}{k_r}&u_k''+u_{ro}u_k'+(u_{ro}+\frac{u_k}{k_r})^2+\frac{2k_r}{k_\phi}(u_{ro}+1)u_\phi \\ \nonumber
&+&(\frac{2(k_\phi^2+k_z^2)}{k_\phi}-2k_\phi)u_\phi'+u_\phi^2+\frac{4}{k_r}u_k+2H_o=0.\label{eq:ukprop}
\eea
These are two non linear ordinary equations for $u_\phi(\xi)$ and $u_k(\xi)$, from which solutions for other velocity components follow as above.  With the present assumptions, the azimuth equation in section (\ref{sect:steadystate}) integrates to give directly the pressure function as is recorded in equations 
(\ref{eq:propeqs}).

The pressure function $H$ is in units of $(\delta\nu_o)^2$ and velocities are in units of $\delta\nu_o$ in these three equations. The constant $H_o$ represents an external pressure.

Without the need for periodic boundary conditions (propeller blades support discontinuities), this system is readily solved numerically by standard routines. Singularities at particular values of $\xi$ do arise, and they must be kept out of the physical range. 

With the assumption $k_r=k_z\zeta$, the independent variable becomes
\be
\xi:=k_\phi \phi+k_r\ln(\delta r)+k_r \label{eq:xiprop}
\ee
which, held constant, defines a logarithmic spiral in the $r$, $\phi$ plane. However we must remember that $dz/dr=\zeta$, so that the logarithmic spiral is actually drawn on cones with exterior angle $arctan(\zeta)$,equal to $\pi/4$ in our example. Because $k_z=1$, we can also put $k_r=\zeta$ in this example.

We can plot a coordinate line in space by writing for constant  $\xi$ the arc length behaviour as a function of $ds$ according to 
\be
\frac{dr}{ds} =\frac{1}{\sqrt{(1+\zeta^2+\frac{k_r^2}{k_\phi^2})}},~~~\frac{d\phi}{ds}=-\frac{k_r}{rk_\phi}\frac{dr}{ds},~~~\frac{dz}{ds}=\zeta\frac{dr}{ds}.\label{eq:coordline}
\ee
These equations are easily integrated to give 
\be
r=\frac{s}{A}+C_r,~~~~\phi=-\frac{k_r}{k_\phi}\ln{\frac{\delta ~s}{A}}+ C_\phi,~~~~z=\frac{\zeta ~s}{A}+C_z,\label{eq:integrations}
\ee
where $A\equiv \sqrt{1+\zeta^2+k_r^2/k_\phi^2}$.
but they are more useful numerically in differential form ($s=0$ is more readily treated).

We will want to compare stream lines of the fluid flow in space with the location of particular logarithmic spiral coordinate lines (on $45^\circ$ cones), considered as possible propeller blades. This is certainly not an engineering exercise, but rather a sketch of where the self-similar equations may be used to help design a log spiral propeller.

One difficulty with these coordinate lines on cones, is that two lines with different constants can arrive at the same point in space for a given set of coordinate differences and different arc lengths. The necessary constraints on the different arc lengths and constant differences ($\Delta C=C_2-C_1$) are 
\be
\Delta C_z=\zeta \Delta C_r,~~~~\frac{s_2}{s_1}=1=\frac{A}{s_1}\Delta C_r~~~~\Delta C_\phi=\frac{k_r}{k_\phi}\ln{\frac{s_2}{s_1}}.\label{eq:intersections}
 \ee
The choices for $\Delta C_r$ and $s_1$ are unconstrained so that there are many possible solutions giving a coordinate singularity like the pole on a sphere.

This is not as serious a singularity as that which may arise in the Navier-Stokes equations themselves, being one that arises in zero pressure fluid flows. Nevertheless an engineering approach would require avoiding this, possibly by experimenting with different constants $k_r$,$k_\phi$, $\zeta$.
We will be content with a one bladed example.

\clearpage
\begin{figure}{}
\begin{tabular}{cc}
\rotatebox{0}{\scalebox{0.35}
{\includegraphics{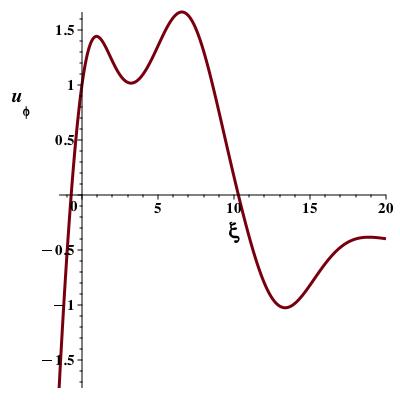}}}&
\rotatebox{0}{\scalebox{0.35}
{\includegraphics{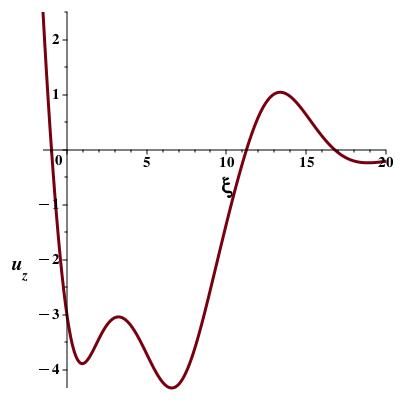}}}\\
\rotatebox{0}{\scalebox{0.35}
{\includegraphics{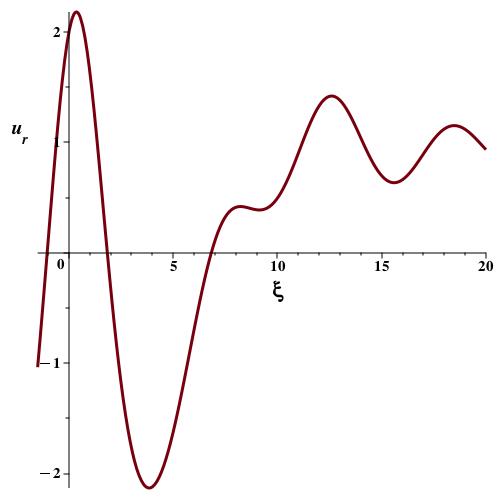}}}&
\rotatebox{0}{\scalebox{0.35}
{\includegraphics{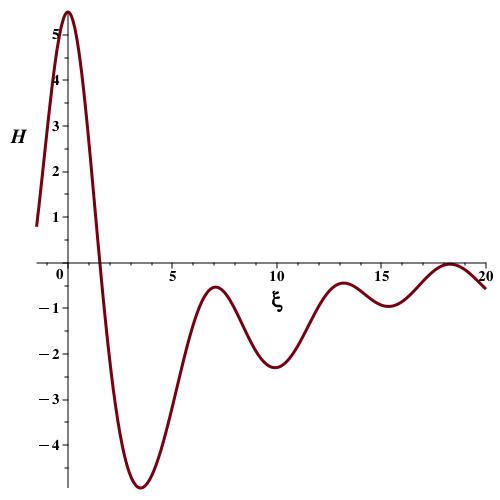}}}
\end{tabular}
\caption{The figure shows the flow velocities and the pressure variation for our standard example with $k_\phi=2$,$k_r=1$ and $\zeta=1$. The initial conditions are $u_k'(0)=u_k(0)=1$ and $u_\phi'(0)=u_\phi(0)=1$. The constant $H_o=0$. There is a singularity in the fluid equations at $\xi=-4.7658$.
   At upper left we show $u_\phi$, $u_z$ at upper right and $u_r$ at lower left. The pressure is at lower right.  \label{fig:Hprop1}}\end{figure}

The flow behaviour shown in figure(\ref{fig:Hprop1}) indicates that the region around $\xi=2..3.5$ is of physical interest. The longitudinal velocity is strongly negative, which would give a positive thrust on a solid blade coinciding with a congruence of coordinate lines. We note that a small range of $\xi(r,\phi,z)$ close to a given value defines a blade surface. The pressure variation would be rather steep across a blade located near $\xi=3.5$. The circular velocity is relatively smooth in this region, and the radial velocity is decelerated rapidly.

\begin{figure}{}
\begin{tabular}{cc}
\rotatebox{0}{\scalebox{0.30}
{\includegraphics{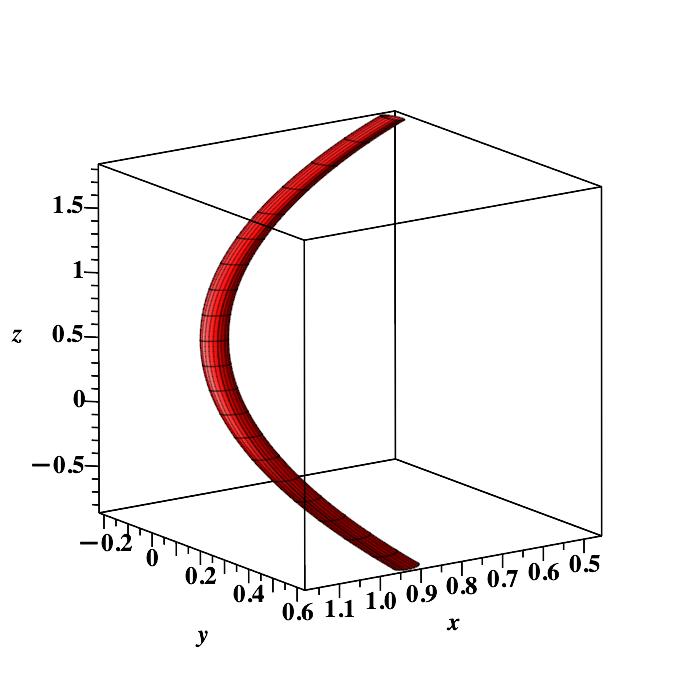}}}&
\rotatebox{0}{\scalebox{0.3}
{\includegraphics{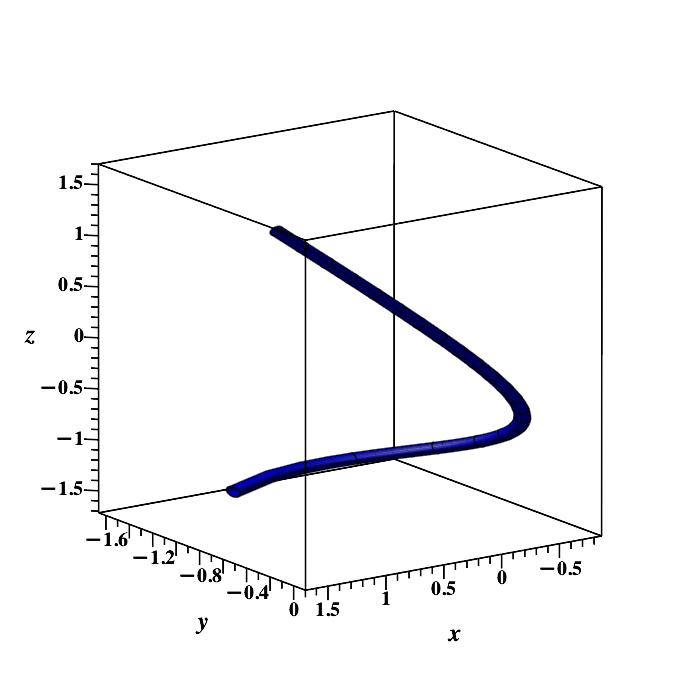}}}\\
\rotatebox{0}{\scalebox{0.3}
{\includegraphics{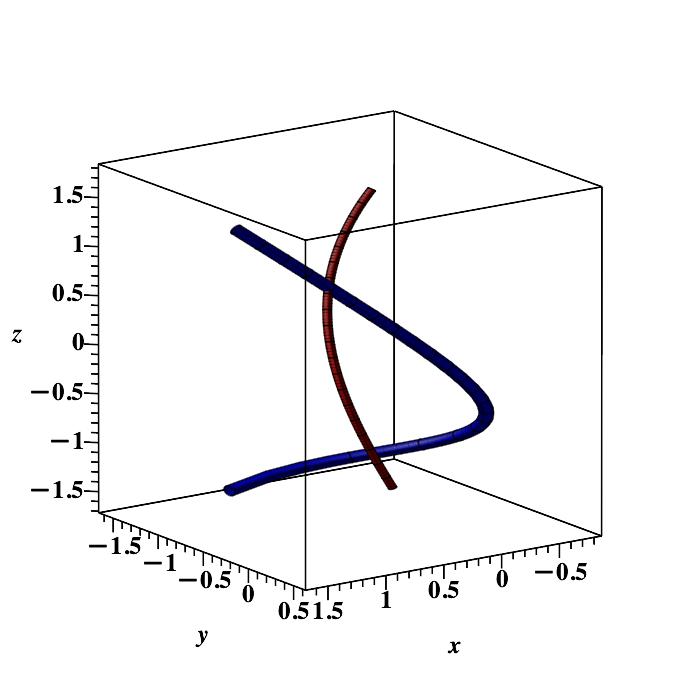}}}&
\rotatebox{0}{\scalebox{0.3}
{\includegraphics{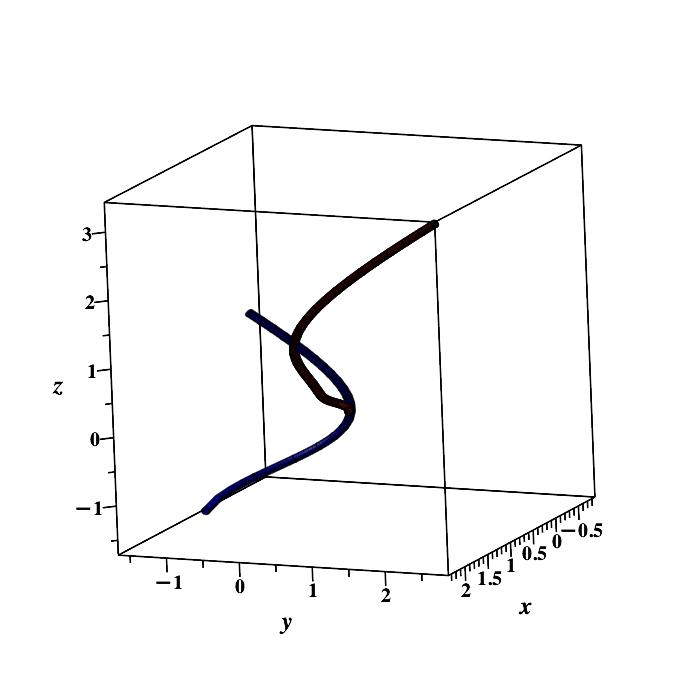}}}
\end{tabular}
\caption{The congruence of stream line shown at upper left passes near or through $r(0)=1$ and $\phi(0)=0$ and $z(0)=1$. It is integrated over $-1\le s\le 2$. At upper right we show the coordinate line line for $\xi=2$, which also passes through $r(0)=1$, $\phi(0)=0$ and $z(0)=0$. At lower left the two are combined as a kind of one bladed logarithmic spiral propeller.
At lower right the same coordinate line is combined with a congruence of stream lines  calculated for $k_\phi=-2$ over the arc length $-2.25\le s\le 4$.\label{fig:Hprop2}}\end{figure}

In figure (\ref{fig:Hprop2}) we show in the upper row a congruence of stream lines in red plus a congruence of coordinate lines in blue near $\xi=2$. The coordinate congruence approximates a logarithmic spiral propeller blade. The pressure function in figure (\ref{fig:Hprop1}) shows $\xi=2$ to be a region of large pressure gradient across such a blade. Figure (\ref{fig:Hprop1}) also shows that the $z$ velocity is strongly negative in this region so that there is thrust along the positive $z$ axis. The radial velocity is more turbulent although locally decelerated, and the azimuthal velocity peaks smoothly.

The lower row of figure (\ref{fig:Hprop2}) combines in space  the coordinate congruence (i.e. blade) with the stream line congruence on the left. The `hub' of the single blade propeller would be near the intersection of the two curves. 

\begin{figure}{}
\begin{tabular}{cc}
\rotatebox{0}{\scalebox{0.35}
{\includegraphics{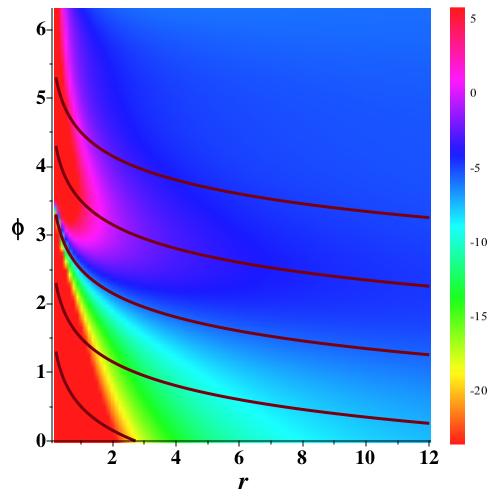}}}&
\rotatebox{0}{\scalebox{0.4}
{\includegraphics{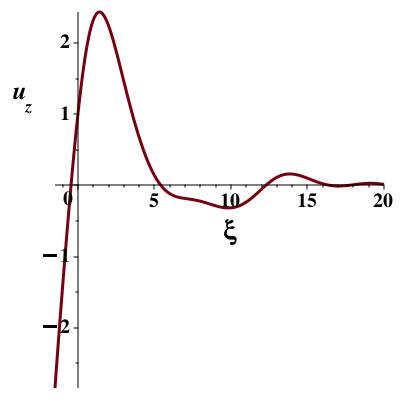}}}
\end{tabular}
\caption{The right hand image shows the $u_z$ velocity for the reversed propeller with $k_\phi=-2$. The flow near $\xi=2$ is now positive, giving reverse thrust. On the left we show $v_z=u_z/\delta r$ for the standard case ($k_\phi=2$) projected onto the $r,\phi$ plane. There is strong negative velocity at small angle and radius. The coordinate lines projected onto the $r,\phi$ plane are shown for $\xi=2$ (the smaller line) until $\xi=10$ in steps of $2$. \label{fig:Hprop3}}\end{figure}

At lower right the same coordinate line is combined with a stream line congruence when the direction of rotation is reversed ($k_\phi=-2$), but all other parameters are the same. The stream line integration in that case is over the range $-4\le s \le 4$ and there is a singularity at $\xi=-6.9615$. In figure (\ref{fig:Hprop3}) we show that the $z$ velocity is also reversed so that there is reverse thrust with the reverse rotation as expected.
 
\clearpage

\clearpage

The left image of figure (\ref{fig:Hprop3}) shows the $v_z=u_z/\delta r$ velocity corresponding to the scaled velocity shown in figure(\ref{fig:Hprop1}). The lines are the coordinate lines stepping from $\xi=2$ to $\xi=10$ in steps of $2$ units. The Navier-Stokes singularity locus would be at $\xi=-4.55$. The right image shows the $u_z$ velocity for the reversed propeller ($k_\phi=-2$). The thrust is reversed. All of the velocities and pressure can be displayed as in the left image, but that is best left for a more engineering study. The singularity in $\xi$ is encountered for the reversed case.

\section{Steady, Non Axially Symmetric, Euler Flow}
\label{sect:Euler}

We restrict our study of the zero viscosity flow to the case $a=2$ and $k_z=0$ and $\zeta=0$  because it allows some comparison with a similar analytic case that includes viscosity in section (\ref{sect:axiallysymmetric}). The parameters $k_r$ and $k_\phi$ are strictly non zero and $\xi=k_\phi\phi+k_r\ln{\delta r}$.

From continuity, we conclude that $u_k=k_\phi u_\phi+k_r u_r$
 is constant, and hence with velocity in units of $u_k$

 \be
 u_r=\frac{1-k_\phi u_\phi}{k_r}.\label{eq:Eur}
 \ee
Moreover the azimuthal Euler equation integrates immediately to give ($H$ in units of $u_k^2$)
\be
H=H_o-\frac{u_\phi}{k_\phi},\label{eq:EH}
\ee
where $H_o$ is an arbitrary constant corresponding to external pressure plus potential. 

The equation for $u_z$ integrates formally to 
\be
u_z=C_2\exp{(\frac{\eta}{k_\phi}-\int\,u_\phi d\eta)},\label{eq:Euz}
\ee
where we have set 
\be
\eta=\frac{k_\phi}{k_r}\xi.\label{eq:eta}
\ee

The last three equations solve the problem once $u_\phi$ is known. The latter follows from the solution of the radial Euler equation which, after substituting for $u_r$ and $H$ from above, takes the form
\be
\frac{du_\phi}{d\eta} =-\frac{1}{k^2}(1+2H_ok_r^2)+\frac{2u_\phi}{k_\phi}-u_\phi^2.\label{eq:Euphi}
\ee
Here the constant $H_o$ is also in units of $u_k^2$, and $k^2=k_r^2+k_\phi^2$. If we set 
\be
A=1+2H_ok_r^2,~~~~B=\frac{k_r}{k}\sqrt{2H_ok_\phi^2-1}, \label{eq:AB}
\ee
then the solution for $u_\phi$ is 
\be
u_\phi=\frac{1}{k_\phi}\big(1-B\,\tan{\big[\frac{B(\eta+C_1)}{k_\phi}\big]}\big).\label{eq:Euphisol}
\ee
This allows all other quantities to be found. The integration constant is $C_1$.

We note that the tangent solution requires that 

\be
2H_o\,k_\phi^2-1\ge 0,\label{eq:singcon}
\ee
which can not happen for $H_o=0$. Otherwise the tangent becomes $i \tanh{[~]}$ and $B\rightarrow iB_2=i(k_r/k)\sqrt{1-2H_o\,k_\phi^2}$ so that with $B_2$ real
\be
u_\phi=\frac{1}{k_\phi}\big(1+B_2\,\tanh{\big[\frac{B_2(\eta+C_1)}{k_\phi}\big]}\big).\label{eq:Euphisol2}
\ee
The condition (\ref{eq:singcon}) that allows the tangent solution, is also the condition that singularities can exist (except at $r=0$ because $v=u/\delta r$). It will require an external pressure $H_ok_\phi^2/u_k^2\ge 1/2$ for a singularity to develop. These will appear where $B(\eta+C_1)/k_\phi=(2n+1)\pi/2$ ($n\ge 0$) or explicitly on a logarithmic spiral (fixed value of $\xi$ and $C_1$) given by 
\be
\frac{1}{k}\sqrt{2H_ok_\phi^2-1}(k_\phi\phi+k_r\ln{\delta r}+C_1)=(2n+1)\frac{\pi}{2}.\label{eq:expsing}
\ee
At $\delta r=1$ the singularity appears at (taking $n=0$)
\be
\phi=\frac{k_r}{k_\phi B}(\pi/2-C_1),
\ee
provided $C_1$ is chosen so that this is less that $2\pi$. With $k_\phi=2$, $k_r=1$ and $H_o=5$, this becomes $\phi=0.28$ on taking $C_1=0$. In radius at fixed angle, say $\phi=0$, the singular spiral with $C_1=0$ is met at $\delta r=1.755$.

The viscous case studied in section (\ref{sect:axiallysymmetric}) had no point singularity. The only difference besides the lack of viscosity is non axial symmetry and a steady flow of this Euler version. 

The quantities $H$ and $u_r$ are given directly by equations (\ref{eq:EH}) and (\ref{eq:Eur}) respectively. Equation (\ref{eq:Euz}) gives for the singular ($B$ real) case 
\be
u_z=C_2\ln\big(\cos{(B(\eta+C_1)/k_\phi)}\big),\label{eq:expuzsol}
\ee
and the $\cos()$ becomes $\cosh()$ when equation (\ref{eq:singcon}) is reversed, and $B$ becomes $B_2$ .

According to equation (\ref{eq:expuzsol}) if $C_2\ne 0$, then $u_z$ has logarithmic singularities at the singularities of $u_\phi$. There are only logarithmic singularities at infinity in the non singular case. 

The radial velocity has singularities where $u_\phi$ does, as does the pressure function. 

This solution adds to the collection of exact solutions that is found in  the book (\cite{DR2006}). Being non axially symmetric and three dimensional it is rather unique. It also contains a singularity (in addition to the divergence at $r=0$) under appropriate boundary conditions and so adds to the the search for these found in (\cite{MB2001}). Detailed properties of the solution 
concerning stability and energy remain to be discovered.

\section{Discussion}

The main contribution of this paper is the reduction of the Navier-Stokes equations, using a combination of invariants under a Lie symmetry as the independent variable. This led to scale invariance and ordinary non-linear equations. The flow is described in terms of logarithmic spirals.

We have used this formulation to study a simple axially symmetric flow, related to astrophysical accretion discs.The flow is free of singularities at all scales. A related Euler flow is also discussed where singularities do occur.

We have also presented two approximations to physical problems. One is Hele-Shaw flow and the  other allows the properties of equiangular propellers to be calculated. The Hele-Shaw example does not allow a rigorous study with our formulation, but the formulation can be of real help in a complete engineering design of logarithmic spiral propeller blades.

\section{Acknowledgements}
Queen's University has supported this work by allowing privileges available to an Emeritus professor. Professor Judith Irwin has provided her usual wise advice.

\section{Appendix A}

In this appendix we illustrate the algebra used in reducing the Navier-Stokes equations to the ordinary equations used in the text. We restrict ourselves to deriving the radial equation of motion in the steady state, because the other equations follow similarly.

We begin by gathering from the text the essential relations. These are ($\zeta=z/r$)
\bea
\delta r=e^{(\delta T)},~~~~~\frac{dT}{dr}&=&e^{-(\delta T)},~~~~~\xi=k_\phi\phi+k_r\delta T+k_o\delta^2\nu_o\tau+k_z \zeta,\nonumber\\
{\bf v}={\bf u}(\xi)e^{((1-a)\delta T)},\bm{\omega}&=&\bm{\varpi}(\xi)e^{(-\alpha T)},\nu=\nu_oe^{((2-a)\delta T)},dt=d\tau~ e^{(\alpha T)}.\label{eq:basics}
\eea
We recall the dynamic equations (\ref{eq:N-S})
\be
\nabla\Psi+\partial_t{\bf v}={\bf v}\times {\bm \omega }-\nu\nabla\times \bm{\omega}.
\ee
These are used together with the incompressible condition $\nabla\cdot {\bf v}=0$. 

 The radial equation follows from the RHS of equation (\ref{eq:N-S}). We need the radial component of the curl of $\bm{\omega}$ and the radial component of ${\bf v}\times \bm{\omega}$. This requires knowing 
$\omega_\phi$ and $\omega_z$. The calculation of $\omega_\phi$ follows from
\be
\omega_\phi=\partial_z v_r-\partial_r v_z=\partial_z(u_r e^{((1-a)\delta T)})-\partial_r(u_z e^{((1-a)\delta T)}),
\ee
which becomes ($\delta r=e^{(\delta T)}$ and $a=\alpha/\delta$)
\be
\omega_\phi=e^{(-a\delta T)}(\delta k_z u_r'-\delta k_r u_z' -\delta(1-a)u_z +\delta k_z \zeta u_z')\equiv \varpi_\phi e^{(-a\delta T)}.
\ee

The component $\omega_z$ follows as
\be
\omega_z=\frac{1}{r}\partial_r(rv_\phi)-\frac{1}{r}\partial_\phi v_r.
\ee
Using the relations (\ref{eq:basics}) at the beginning of this appendix, this becomes
\be
\omega_z=\delta e^{(-a\delta T)}((2-a)u_\phi+k_ru_\phi'-k_\phi u_r'-k_z\zeta u_\phi')\equiv \varpi_ze^{(-a\delta T)}.
\ee

We can now compute the radial component of the curl of the vorticity
as multiplied by $\nu$ from 
\be
\nu(\nabla\times \bm{\omega})_r=\nu_o e^{((1-2a)\delta T)}\big(\frac{1}{r}(\partial_\phi\omega_z)-\partial_z \omega_{\phi}\big).
\ee
Using the relations (\ref{eq:basics}) at the beginning of the appendix, and the expression for $\omega_z$ and $\omega_\phi$ above we find
\bea
&\nu&(\nabla\times \omega)_r=\nu_o e^{((1-2a)\delta T)}\delta^2 \big( k_\phi((2-a)u_\phi'+(k_r-k_z\zeta) u_\phi''-k_\phi u_r'')\nonumber\\
&-&k_z(a u_z'-k_r u_z''+k_z u_r''+\zeta k_z u_z'')\big).
\eea

The second term on the right of the radial component of equation (\ref{eq:N-S}), is an inertial quantity equal to the radial component of the cross product of the velocity and the vorticity in the form
\be
({\bf v}\times \bm{\omega})_r=v_\phi\omega_z-v_z\omega_\phi,
\ee
which requires only substitution from the results above to take the form 
\bea
({\bf v}\times \bm{\omega})_r&=&e^{((1-2a)\delta T)}\big[\delta u_\phi((2-a)u_\phi+(k_r-\zeta k_z) u_\phi'-k_\phi u_r')\nonumber\\
&+&\delta u_z((k_r-k_z\zeta) u_z'+(1-a)u_z-k_z u_r')\big].
\eea

By subtraction of these two terms on the right hand side (RHS) of equation (\ref{eq:N-S}), and subsequent equality to the left hand side (LHS) computed below, we can obtain the radial equation.  

 On the LHS
 \be
 \partial_r(\Psi)=\partial_r((H+\frac{{\bf u}^2}{2})e^{(2(1-a)\delta T)}),
 \ee
 which becomes
 \be
 \partial_r\Psi=e^{((1-2a)\delta T)}\big[2\delta(1-a)(H+\frac{{\bf u}^2}{2})+\delta (k_r-k_z\zeta)(H'+u_r u_r'+u_\phi u_\phi'+u_z u_z')\big].
 \ee

Finally, we equate the terms left and right, introduce $u_k$, and rearrange slightly, to obtain equation (\ref{eq:NSradial}).

\bibliography{newfluidpapernew}

\providecommand{\noopsort}[1]{}\providecommand{\singleletter}[1]{#1}%
\begin{thebibliography}{12}%
\makeatletter
\providecommand \@ifxundefined [1]{%
 \@ifx{#1\undefined}
}%
\providecommand \@ifnum [1]{%
 \ifnum #1\expandafter \@firstoftwo
 \else \expandafter \@secondoftwo
 \fi
}%
\providecommand \@ifx [1]{%
 \ifx #1\expandafter \@firstoftwo
 \else \expandafter \@secondoftwo
 \fi
}%
\providecommand \natexlab [1]{#1}%
\providecommand \enquote  [1]{``#1''}%
\providecommand \bibnamefont  [1]{#1}%
\providecommand \bibfnamefont [1]{#1}%
\providecommand \citenamefont [1]{#1}%
\providecommand \href@noop [0]{\@secondoftwo}%
\providecommand \href [0]{\begingroup \@sanitize@url \@href}%
\providecommand \@href[1]{\@@startlink{#1}\@@href}%
\providecommand \@@href[1]{\endgroup#1\@@endlink}%
\providecommand \@sanitize@url [0]{\catcode `\\12\catcode `\$12\catcode
  `\&12\catcode `\#12\catcode `\^12\catcode `\_12\catcode `\%12\relax}%
\providecommand \@@startlink[1]{}%
\providecommand \@@endlink[0]{}%
\providecommand \url  [0]{\begingroup\@sanitize@url \@url }%
\providecommand \@url [1]{\endgroup\@href {#1}{\urlprefix }}%
\providecommand \urlprefix  [0]{URL }%
\providecommand \Eprint [0]{\href }%
\providecommand \doibase [0]{http://dx.doi.org/}%
\providecommand \selectlanguage [0]{\@gobble}%
\providecommand \bibinfo  [0]{\@secondoftwo}%
\providecommand \bibfield  [0]{\@secondoftwo}%
\providecommand \translation [1]{[#1]}%
\providecommand \BibitemOpen [0]{}%
\providecommand \bibitemStop [0]{}%
\providecommand \bibitemNoStop [0]{.\EOS\space}%
\providecommand \EOS [0]{\spacefactor3000\relax}%
\providecommand \BibitemShut  [1]{\csname bibitem#1\endcsname}%
\let\auto@bib@innerbib\@empty
\bibitem [{\citenamefont {{Meneveau}}\ and\ \citenamefont
  {{Katz}}(2000)}]{MK2000}%
  \BibitemOpen
  \bibfield  {author} {\bibinfo {author} {\bibfnamefont {C.}~\bibnamefont
  {{Meneveau}}}\ and\ \bibinfo {author} {\bibfnamefont {J.}~\bibnamefont
  {{Katz}}},\ }\bibfield  {title} {\enquote {\bibinfo {title}
  {{Scale-Invariance and Turbulence Models for Large-Eddy Simulation}},}\
  }\href {\doibase 10.1146/annurev.fluid.32.1.1} {\bibfield  {journal}
  {\bibinfo  {journal} {Annual Review of Fluid Mechanics}\ }\textbf {\bibinfo
  {volume} {32}},\ \bibinfo {pages} {1--32} (\bibinfo {year}
  {2000})}\BibitemShut {NoStop}%
\bibitem [{\citenamefont {{Schaefer-Rolffs}}(2019)}]{USR2019}%
  \BibitemOpen
  \bibfield  {author} {\bibinfo {author} {\bibfnamefont {U.}~\bibnamefont
  {{Schaefer-Rolffs}}},\ }\bibfield  {title} {\enquote {\bibinfo {title} {{The
  Scale Invariance Criterion for Geophysical Fluids}},}\ }\href@noop {}
  {\bibfield  {journal} {\bibinfo  {journal} {European Journal of Mechanics-B
  fluids}\ }\textbf {\bibinfo {volume} {74}},\ \bibinfo {pages} {92--98}
  (\bibinfo {year} {2019})}\BibitemShut {NoStop}%
\bibitem [{\citenamefont {{Carter}}\ and\ \citenamefont
  {{Henriksen}}(1991)}]{CH1991}%
  \BibitemOpen
  \bibfield  {author} {\bibinfo {author} {\bibfnamefont {B.}~\bibnamefont
  {{Carter}}}\ and\ \bibinfo {author} {\bibfnamefont {R.~N.}\ \bibnamefont
  {{Henriksen}}},\ }\bibfield  {title} {\enquote {\bibinfo {title} {{A
  systematic approach to self-similarity in Newtonian space-time}},}\ }\href
  {\doibase 10.1063/1.529103} {\bibfield  {journal} {\bibinfo  {journal}
  {Journal of Mathematical Physics}\ }\textbf {\bibinfo {volume} {32}},\
  \bibinfo {pages} {2580--2597} (\bibinfo {year} {1991})}\BibitemShut {NoStop}%
\bibitem [{\citenamefont {Henriksen}(2015)}]{H2015}%
  \BibitemOpen
  \bibfield  {author} {\bibinfo {author} {\bibfnamefont {R.}~\bibnamefont
  {Henriksen}},\ }\href@noop {} {\emph {\bibinfo {title} {Scale Invariance}}}\
  (\bibinfo  {publisher} {Wiley-VCH},\ \bibinfo {year} {2015})\BibitemShut
  {NoStop}%
\bibitem [{\citenamefont {Drazin}\ and\ \citenamefont {Riley}(2006)}]{DR2006}%
  \BibitemOpen
  \bibfield  {author} {\bibinfo {author} {\bibfnamefont {P.~G.}\ \bibnamefont
  {Drazin}}\ and\ \bibinfo {author} {\bibfnamefont {N.}~\bibnamefont {Riley}},\
  }\href@noop {} {\emph {\bibinfo {title} {The Navier-Stokes Equations: A
  Classification of Flows and Exact Solutions}}},\ London Mathematical Society
  Lecture Note Series\ (\bibinfo  {publisher} {Cambridge University Press},\
  \bibinfo {year} {2006})\BibitemShut {NoStop}%
\bibitem [{\citenamefont {{Majda}}\ and\ \citenamefont
  {{Bertozzi}}(2001)}]{MB2001}%
  \BibitemOpen
  \bibfield  {author} {\bibinfo {author} {\bibfnamefont {A.~J.}\ \bibnamefont
  {{Majda}}}\ and\ \bibinfo {author} {\bibfnamefont {A.~L.}\ \bibnamefont
  {{Bertozzi}}},\ }\href@noop {} {\emph {\bibinfo {title} {{Vorticity and
  Incompressible Flow}}}}\ (\bibinfo  {publisher} {Cambridge University
  Press},\ \bibinfo {year} {2001})\BibitemShut {NoStop}%
\bibitem [{\citenamefont {Wang}\ \emph {et~al.}(2024)\citenamefont {Wang},
  \citenamefont {Sprinkle}, \citenamefont {Zuo},\ and\ \citenamefont
  {Ristroph}}]{WSZR2024}%
  \BibitemOpen
  \bibfield  {author} {\bibinfo {author} {\bibfnamefont {K.}~\bibnamefont
  {Wang}}, \bibinfo {author} {\bibfnamefont {B.}~\bibnamefont {Sprinkle}},
  \bibinfo {author} {\bibfnamefont {M.}~\bibnamefont {Zuo}}, \ and\ \bibinfo
  {author} {\bibfnamefont {L.}~\bibnamefont {Ristroph}},\ }\bibfield  {title}
  {\enquote {\bibinfo {title} {Centrifugal flows drive reverse rotation of
  feynman's sprinkler},}\ }\href {\doibase 10.1103/PhysRevLett.132.044003}
  {\bibfield  {journal} {\bibinfo  {journal} {Phys. Rev. Lett.}\ }\textbf
  {\bibinfo {volume} {132}},\ \bibinfo {pages} {044003} (\bibinfo {year}
  {2024})}\BibitemShut {NoStop}%
\bibitem [{Note1()}]{Note1}%
  \BibitemOpen
  \bibinfo {note} {Here the co-vector ${\protect \bf d}$ is resolved along
  space-time units so for the length co-vector, there is one unit of space and
  zero units of time}\BibitemShut {NoStop}%
\bibitem [{\citenamefont {Frisch}(1995)}]{F1995}%
  \BibitemOpen
  \bibfield  {author} {\bibinfo {author} {\bibfnamefont {U.}~\bibnamefont
  {Frisch}},\ }\href@noop {} {\emph {\bibinfo {title} {Turbulence: The Legacy
  of A. N. Kolmogorov}}}\ (\bibinfo  {publisher} {Cambridge University Press},\
  \bibinfo {year} {1995})\BibitemShut {NoStop}%
\bibitem [{\citenamefont {{Lynden-Bell}}\ and\ \citenamefont
  {{Pringle}}(1974)}]{LBP1974}%
  \BibitemOpen
  \bibfield  {author} {\bibinfo {author} {\bibfnamefont {D.}~\bibnamefont
  {{Lynden-Bell}}}\ and\ \bibinfo {author} {\bibfnamefont {J.~E.}\ \bibnamefont
  {{Pringle}}},\ }\bibfield  {title} {\enquote {\bibinfo {title} {{The
  evolution of viscous discs and the origin of the nebular variables.}}}\
  }\href {\doibase 10.1093/mnras/168.3.603} {\bibfield  {journal} {\bibinfo
  {journal} {mnras}\ }\textbf {\bibinfo {volume} {168}},\ \bibinfo {pages}
  {603--637} (\bibinfo {year} {1974})}\BibitemShut {NoStop}%
\bibitem [{\citenamefont {{Abramowitz}}\ and\ \citenamefont
  {{Stegun}}(1970)}]{abst1970}%
  \BibitemOpen
  \bibfield  {author} {\bibinfo {author} {\bibfnamefont {M.}~\bibnamefont
  {{Abramowitz}}}\ and\ \bibinfo {author} {\bibfnamefont {I.~A.}\ \bibnamefont
  {{Stegun}}},\ }\href@noop {} {\emph {\bibinfo {title} {{Handbook of
  mathematical functions : with formulas, graphs, and mathematical tables}}}}\
  (\bibinfo  {publisher} {National Bureau of Standards, Applied Mathematics
  Series 55},\ \bibinfo {year} {1970})\BibitemShut {NoStop}%
\bibitem [{\citenamefont {{Gowen}}, \citenamefont {{Videb{\ae}k}},\ and\
  \citenamefont {{Nagel}}(2024)}]{GVN2024}%
  \BibitemOpen
  \bibfield  {author} {\bibinfo {author} {\bibfnamefont {S.~D.}\ \bibnamefont
  {{Gowen}}}, \bibinfo {author} {\bibfnamefont {T.~E.}\ \bibnamefont
  {{Videb{\ae}k}}}, \ and\ \bibinfo {author} {\bibfnamefont {S.~R.}\
  \bibnamefont {{Nagel}}},\ }\bibfield  {title} {\enquote {\bibinfo {title}
  {{Measurement of pressure gradients near the interface in the viscous
  fingering instability}},}\ }\href {\doibase 10.1103/PhysRevFluids.9.053902}
  {\bibfield  {journal} {\bibinfo  {journal} {Physical Review Fluids}\ }\textbf
  {\bibinfo {volume} {9}},\ \bibinfo {eid} {053902} (\bibinfo {year} {2024})},\
  \Eprint {http://arxiv.org/abs/2402.01924} {arXiv:2402.01924
  [physics.flu-dyn]} \BibitemShut {NoStop}%
\end{thebibliography}%
 \end{document}